     \def\lsim{\mbox{\hspace{1ex}}
               \hbox{\raisebox{.4ex}{$<$}}\mbox{\hspace{-1.7ex}}
              {\lower.8ex \hbox{$\sim$}}\mbox{\hspace{1.1ex}}}
     \def\ssim{\mbox{\hspace{1ex}}
               \hbox{\raisebox{.4ex}{$>$}}\mbox{\hspace{-1.7ex}}
              {\lower.8ex \hbox{$\sim$}}\mbox{\hspace{1.1ex}}}

     \def\@cite#1#2{\leavevmode\hbox{$^{\mbox{\the\scriptfont0 #1}}$}}

\documentstyle[12pt]{article}

      1
     \font\elevenrm=cmr10 scaled\magstep 1
     \font\elevenit=cmti10 scaled\magstep 1

     \renewenvironment{thebibliography}[1]
      { \elevenrm
        \begin{list}{\arabic{enumi}.}
         {\usecounter{enumi} \setlength{\parsep}{0pt}
          \setlength{\itemsep}{3pt} \settowidth{\labelwidth}{#1.}
          \sloppy
         }}{\end{list}}

\begin{document}

\begin{flushright}
\begin{tabular}{l}
HUPD-9314\hspace{0.5cm}\\
May 1993
\end{tabular}
\end{flushright}
\vglue 0.9cm
\begin{center}
{{\Huge Nambu-Jona-Lasinio Model\\
 in Curved Space-Time
\footnote{Work supported in part by the Monbusho Grant-in-Aid for
 Scientific Research (C)
 No. 04640301 and the Monbusho Grant-in-Aid for Encouragement of
 Young Scientists No. 92011.
}
\\}
\vglue 1.5cm
{\Large T.~Inagaki, T.~Muta, and S.~D.~Odintsov
\footnote{On leave of absence from Tomsk Pedagogical Institute,
 634041 Tomsk, Russia.
}
\\}
\baselineskip=13pt
\vglue 0.5cm
{\elevenit Department of Physics, Hiroshima University\\}
\baselineskip=12pt
\vglue 0.15cm
{\elevenit Higashi-Hiroshima, Hiroshima 724, Japan\\}
\vglue 4cm

{\Large ABSTRACT}}
\end{center}
\vglue 0.6cm
{\rightskip=3pc
 \leftskip=3pc
 \elevenrm\baselineskip=18pt
 \noindent
 The phase structure of Nambu-Jona-Lasinio model with N-component fermions
 in curved space-time is studied in the leading order of the 1/N expansion.
 The effective potential for composite operator $\bar{\psi}\psi$ is calculated
 by using the normal coordinate expansion in the Schwinger proper-time method.
 The existence of the first-order phase transition caused by the change of
 the space-time curvature is confirmed and the dynamical mass of the fermion
 is calculated as a simultaneous function of the curvature and the
 four-fermion coupling constant.
 The phase diagram in the curvature and the coupling constant is obtained.
\vglue 0.5cm}

\newpage

\baselineskip=20pt
\noindent
It is quite interesting to make investigations of how the phase transition
 in quantum field theory takes place under the circumstance of the early
 universe.
 In paticular much interest has been taken in clarifying the mechanism of
 the spontaneous symmetry breaking in the grand unified theories under the
 influence of the temperature, density and external gravity\cite{EF}.\\
\hspace*{\parindent}
In the standard grand unified theories the Higgs field appears as an
 elementary field and has a Yukawa coupling to fermion fields giving masses
 to the fermions.
 As in the technicolor model for electroweak theory it is possible to assume
 that the Higgs field may be a composite system of some fundamental fermion
 fields.
 We shall call the model based on such an assumption generically the
 composite Higgs model.\\
\hspace*{\parindent}
As far as we know a very little number of works have been reported in the
 study of the composite Higgs models under the circumstance of the early
 universe\cite{RG}.
 Therefore we launched our plan to make a systematic study of the composite
 Higgs models under the circumstance of the early universe.
 Before going into a direct investigation of the composite Higgs models it
 may be better to start with a fundamental study of prototype models of
 the composite Higgs particle with
 finite temperature, finite density and external gravity.
 In the present communication we take Nambu-Jona-Lasinio model as one of
 such prototype models and make a brief report of our work in the study
 of the phase structure of Nambu-Jona-Lasinio model in external gravity.\\
\hspace*{\parindent}
The Nambu-Jona-Lasinio model\cite{NJL} in curved space-time is defined by
 the action,
\begin{equation}
     S = \int d^{4}x \sqrt{-g}
              \left[
              \bar{\psi}i\gamma^{\mu}(x)\nabla_{\mu}\psi
              +\frac{\lambda}{2\mbox{N}}
                    \left\{
                    (\bar{\psi}\psi)^{2}+(\bar{\psi}i\gamma_{5}\psi)^{2}
                    \right\}
              \right]\: ,
\end{equation}
\noindent
where $g$ is the determinant of the space-time metric $g_{\mu\nu}$,
 $\gamma_{\mu}(x)$ the Dirac matrix in curved space-time, $\nabla_{\mu}\psi$
 the covariant derivative of the fermion field $\psi$, N the number of the
 fermion species.
 We work in the scheme of the 1/N expansion and perform our calculation in
 the leading order of the expansion.
 Our notation is basically in conformity with the $(- - -)$ convention
 in the book by Misner, Thorne and Wheeler\cite{GR}.\\
     In practical calculations it is more convenient to introduce auxiliary
 fields $\sigma (x)$ and $\pi(x)$ and start with the following action
 instead of Eq. $(1)$,
\begin{equation}
     S = \int d^{4}x\sqrt{-g}\left[\bar{\psi}i\gamma^{\mu}(x)
                     \nabla_{\mu} \psi
                    -\frac{N}{2\lambda}(\sigma^{2}+\pi^{2})
                    - \bar{\psi}(\sigma+i\gamma_{5}\pi)\psi
                    \right]\: .
\end{equation}
\noindent
\hspace*{\parindent}
In the flat space-time it is well-known that the global abelian chiral
 symmetry posessed by the Lagrangian $(1)$ and $(2)$ is broken spontaneously
 if the coupling constant $\lambda$ exceeds a critical value.
 Our purpose in the present paper is to see whether this phenomenon is
 modified under the influence of external gravity.
 We start with the generating functional given by
\begin{equation}
     Z[\eta, \bar{\eta}] = \int D\psi D\bar{\psi} D\sigma D\pi \exp
              \left\{
              iS+i\bar{\eta}\psi+i\bar{\psi}\eta
              \right\}\: ,
\end{equation}
\noindent
where $\eta$ and $\bar{\eta}$ are source functions. Integrating over the
 fermion fields with $\eta=\bar{\eta}=0$ we find
\begin{equation}
     Z[0,0] = \int D\sigma D\pi \exp \{i \mbox{N} S_{eff}\}\: ,
\end{equation}
\noindent
where we pulled out an obvious factor N in defining the semiclassical
 effective action $S_{eff}$:
\begin{equation}
     S_{eff} = \int d^{4}x \sqrt{-g}
                      \left\{
                      -\frac{1}{2\lambda}(\sigma^{2}+\pi^{2})
                      \right\}
              -i\ln \mbox{Det}
                      \left\{
                      i\gamma^{\mu}(x)\nabla_{\mu}
                      -(\sigma+i\gamma_{5}\pi)
                      \right\} \: .
\end{equation}
\noindent
In the leading order of the 1/N expansion the effective action $\Gamma$
 (with N factored out) is just equal to $S_{eff}$. Thus we obtain
\begin{equation}
  \Gamma[\sigma, \pi] = S_{eff}[\sigma, \pi] + \mbox{O}(1/\mbox{N})\: .
\end{equation}
The effective potential (with N factored out) in the leading order of the
 1/N expansion is then given by
\begin{equation}
     V(\sigma, \pi) = \frac{1}{2\lambda}(\sigma^{2}+\pi^{2})
                      +i\mbox{Tr}\ln
                      \langle x |\{
                      i\gamma^{\mu}(x)\nabla_{\mu}
                      -(\sigma+i\gamma_{5}\pi)\}
                      |x \rangle \: .
\end{equation}
\noindent
In Eq. $(7)$ the variables $\sigma$ and $\pi$ are regarded as constant.\\
\hspace*{\parindent}
To estimate the second term in the right-hand side of Eq. $(7)$ we adopt
 the Schwinger proper time method\cite{SPT}. For this purpose we rewrite
 Eq. $(7)$ in the following form,
\begin{equation}
  V(\sigma, \pi) = \left.\frac{1}{2\lambda}(\sigma^{2}+\pi^{2})
                 -i\mbox{Tr}\ln S(x, x; s)\right|_{s=\sigma+i\gamma_{5}\pi}
                 \: ,
\end{equation}
\noindent
where the Green function defined by
\begin{equation}
  S(x, y; s) = \langle x|(i\gamma^{\mu}\nabla_{\mu}-s)^{-1}|y\rangle
\end{equation}
is the solution of the equation
\begin{equation}
     \left\{
     i\gamma^{\mu}(x)\nabla_{\mu}-s
     \right\}S(x,y;s)=\frac{1}{\sqrt{-g(x)}}\delta^{4}(x-y)\: .
\end{equation}
\noindent
We would like to calculate the Green function $S(x, y, s)$ in the
 approximation of weakly varying gravity where we neglect any terms
 involving derivatives of the metric tensor higher than the third derivative.
 We use the Riemann normal coordinate expansion\cite{RNC}.
 Here we keep only terms independent of curvature $R$ and terms linear in $R$.
 The calculation proceeds just in parallel with the one given in the paper
 by Parker and Toms\cite{PT}. The result is
\begin{eqnarray}
    S(x,x;s) & = &
                   \int \frac{d^{4}k}{(2\pi)^{4}}
                   \left[(\gamma^{\hat{a}}k_{\hat{a}}+s)
                   \frac{1}{k^{2}-s^{2}}
                   -\frac{1}{12}R(\gamma^{\hat{a}}k_{\hat{a}}+s)
                   \frac{1}{(k^{2}-s^{2})^{2}}\right.           \nonumber \\
             &   & \left. +\frac{2}{3}R_{\mu\nu}k^{\mu}k^{\nu}
                   (\gamma^{\hat{a}}k_{\hat{a}}+s)
                   \frac{1}{(k^{2}-s^{2})^{3}}
                   -\frac{1}{2}\gamma^{\hat{a}}{\cal J}^{\hat{c}\hat{d}}
                   R_{\hat{c}\hat{d}\hat{a}\mu}k^{\mu}
                   \frac{1}{(k^{2}-s^{2})^{2}}\right]\: ,       \nonumber \\
\end{eqnarray}
\noindent
where
\begin{equation}
{\cal J}^{\hat{a}\hat{b}}=\frac{1}{4}[\gamma^{\hat{a}},\gamma^{\hat{b}}] \: ,
\end{equation}
and Latin indices with a caret symbol are vierbein indices.
 It may be seen from Eqs. $(8)$ and $(11)$ that $V(\sigma, \pi)$ is
 symmetric in $\sigma$ and $\pi$ and so it is enough to discuss $V(\sigma, 0)$
 insted of the full expression in the following arguments.
 Inserting Eq. $(11)$ in Eq. $(8)$ and making the Fourier transformation
 we obtain
\begin{eqnarray}
    V(\sigma ,0)
             & = & \frac{1}{2\lambda}\sigma^{2}                 \nonumber \\
             &   & -i\mbox{Tr}\int_{0}^{\sigma}ds
                   \int \frac{d^{4}k}{(2\pi)^{4}}
                   \left[(\gamma^{\hat{a}}k_{\hat{a}}+s)
                   \frac{1}{k^{2}-s^{2}}
                   -\frac{1}{12}R(\gamma^{\hat{a}}k_{\hat{a}}+s)
                   \frac{1}{(k^{2}-s^{2})^{2}}\right.           \nonumber \\
             &   & \left. +\frac{2}{3}R_{\mu\nu}k^{\mu}k^{\nu}
                   (\gamma^{\hat{a}}k_{\hat{a}}+s)
                   \frac{1}{(k^{2}-s^{2})^{3}}
                   -\frac{1}{2}\gamma^{\hat{a}}{\cal J}^{\hat{c}\hat{d}}
                   R_{\hat{c}\hat{d}\hat{a}\mu}k^{\mu}
                   \frac{1}{(k^{2}-s^{2})^{2}}\right] \: .      \nonumber \\
\end{eqnarray}
\noindent
To perform the momentum integration we make the Wick rotation and regularize
 the divergent integral by the cut-off method.
 We derive the following expression for the effective potential,
\newpage
\begin{eqnarray}
    V(\sigma, 0)
             & = & V(0, 0)
                   +\frac{1}{2\lambda}{\sigma}^{2}             \nonumber \\
             &   & -\frac{1}{(4\pi^{2})}
                   \left[{\sigma}^{2}\Lambda^{2}
                   +\Lambda^{4}
                   \ln \left(1+\frac{{\sigma}^{2}}{\Lambda^{2}}\right)
                   -{\sigma}^{4}
                   \ln \left(1+\frac{\Lambda^{2}}{{\sigma}^{2}}\right)
                   \right]                                      \nonumber \\
             &   & -\frac{1}{(4\pi)^{2}}\frac{R}{6}
                   \left[-{\sigma}^{2}
                   \ln \left(1+\frac{\Lambda^{2}}{{\sigma}^{2}}\right)
                   +\frac{\Lambda^{2}{\sigma}^{2}}{\Lambda^{2}+{\sigma}^{2}}
                   \right] \: .
\end{eqnarray}
\noindent
\hspace*{\parindent}
In Fig. $1$ a typical behavior of the effective potential $V(\sigma, 0)$
 as a function of curvature $R$ is given for fixed four-fermion coupling
 constant $\lambda$.
 As is well-known in the flat space-time, the chiral symmetry is broken
 if $\lambda > \lambda_{0}$ with
\begin{equation}
     \lambda_{0}=\frac{4\pi^{2}}{\Lambda^{2}} \: .
\end{equation}
\noindent
In drawing Fig. $1$ the coupling constant $\lambda$ is kept in the region
 $\lambda > \lambda_{0}$.
 It is clearly seen in Fig. $1$ that the first-order phase transition
 takes place as $R$ changes.
 Note, however, that the transition is found to be of second order for
 $\lambda \leq \lambda_{0}$ .
     The dynamical mass of the fermion is calculated by analyzing the gap
 equation $\partial V(\sigma, 0)/\partial\sigma = 0$ :
\begin{eqnarray}
    \left.\frac{\partial V(\sigma, 0)}{\partial \sigma}
    \right|_{\sigma=\sigma_{0}}
             & = & \frac{\sigma_{0}}{\lambda}-\frac{\sigma_{0}}{4\pi^{2}}
                   \left[\Lambda^{2}
                   -\sigma_{0}^{2}
                   \ln \left(1+\frac{\Lambda^{2}}{\sigma_{0}^{2}}\right)
                   \right]                                      \nonumber \\
             &   & -\frac{\sigma_{0}}{48\pi^{2}}R
                   \left[
                   -\ln \left(1+\frac{\Lambda^{2}}{\sigma_{0}^{2}}\right)
                   +\frac{\Lambda^{2}}{\Lambda^{2}+\sigma_{0}^{2}}
                   +\frac{\Lambda^{4}}{(\Lambda^{2}+\sigma_{0}^{2})^{2}}
                   \right]                                      \nonumber \\
             & = & 0 \: .
\end{eqnarray}
\noindent
The solution $\sigma_{0}$ of the gap equation $(16)$ corresponds to the vacuum
 expectation value of the composite field $\bar{\psi}\psi$ in the true vacuum
 and is equal to the dynamical mass of the fermion, $m=\sigma_{0}$.
In Fig. $2$ the dynamical mass of the fermion is plotted as a function of the
 curvature $R$ for fixed $\lambda$.
 The coupling constant $\lambda$ is kept again in the range
 $\lambda > \lambda_{0}$.
 The behavior of the dynamical mass as shown in Fig. $2$ is characteristic
 for the relatively small coupling constant,
 $\lambda_{0} < \lambda \lsim 2\lambda_{0}$.
 For larger coupling, $\lambda \ssim 2\lambda_{0}$, the behavior near the
 critical point $R=R_{cr}$ is quite different from the one in Fig. $2$:
 the curve representing the dynamical mass is bent upward near $R=R_{cr}$.
 At any rate there is observed the gap in the dynamical mass at the critical
 point $R=R_{cr}$ reflecting the nature of the first-order phase transition.
 By the direct numerical analysis we find that there is no gap at
 $R=R_{cr}=0$ if $\lambda \leq \lambda_{0}$ and so the phase
 transition is of second order.\\
\hspace*{\parindent}
 It is possible to obtain the critical value of the curvature $R$ and the
 coupling constant $\lambda$ by observing the behavior of the effective
 potential.
 The critical values $R_{cr}$ and $\lambda_{cr}$ constitute a critical
 curve in the $R-\lambda$ plane as shown in Fig. $3$.
 It is found from Fig. $3$ that for large positive $R$ the chiral symmetry
 is restored even if $\lambda$ is kept in the region of the broken phase
 for $R = 0$.
 On the other hand the chiral symmetry is always broken for negative $R$
 irrespective of the value of $\lambda$.
 This latter conclusion comes from the particular behavior of the effective
 potential $(14)$ with $\pi=0$, i. e., the term proportional to $R$
 in Eq. $(14)$ with $\pi=0$ dominates over other terms if $R<0$ and makes
 the point $\sigma=0$ an unstable stationary point.
 The above peculiar behavior of our effective potential may be due to our
 approximation to keep only the first-order term in $R$ in the normal
 coordinate expansion and should be left for the future detailed study.\\
\hspace*{\parindent}
In the above discussions concerning Figs. $1$, $2$ and $3$ we dealt with the
 relatively large value of the curvature, e. g., in Fig. $1$
 $R_{cr}/\Lambda^{2}=0.656$ which is large compared with
 $\sigma_{0} /\Lambda=0.2\sim 0.3$ .
 Since we work in the normal coordinate
 expansion and keep only terms up to the one linear in $R$, large
 $R_{cr}/\Lambda^{2}$ compared with $\sigma /\Lambda$ may spoil the validity
 of the expansion.
 One should, however, note that the terms quadratic or higher in $R$ are
 not divergent by dimensional reason and the coefficient of $R^{2}$ does
 not carry logarithmic functions of $\sigma^{2}/\Lambda^{2}$.
 Hence the $R^{2}$ term is relatively suppressed compared with the term
 linear in $R$.
 Accordingly relatively large $R_{cr}/\Lambda^{2}$ may be within the scope
of our approximation.\\
\hspace*{\parindent}
In the present model we found within our approximation that the first-order
 phase transition takes place as the curvature varies and the fermion mass
 is generated when the curvature is smaller than the critical value.
 Our conclusion has some similarity with the previous results in the massless
 scalar theory with $\phi^{4}$ coupling\cite{SH}.
 The four-fermion theory in 2 dimensions (Gross-Neveu model\cite{GN})
 has been investigated in curved space-time\cite{GNC} and there
 the second-order phase transition is observed in contrast with our result.

\vglue 0.6cm

\subsection*{REFERENCES}
\vglue 0.1cm

\vglue 0.5cm

\newpage

\subsection*{FIGURE CAPTIONS}
\vglue 0.1cm
Fig. $1$. The typical behavior of the effective potential $V$ is shown
 for fixed $\lambda$\\
\hspace*{\parindent} $(= 1.25\lambda_{0}, \lambda_{0}=4\pi^{2}/\Lambda^{2})$
 as a function of the curvature $R$ where\\
\hspace*{\parindent}
 $R_{cr}=0.656\Lambda^{2}$.
\vglue 0.8cm
\noindent
Fig. $2$. The solution $\sigma_{0}$ of the gap equation $(17)$ is shown as a
 function of the\\
\hspace*{\parindent}curvature $R$ where $\lambda=1.25\lambda_{0}$ and
 $R_{cr}=0.656\Lambda^{2}$.
\vglue 0.8cm
\noindent
Fig. $3$. The phase diagram in $\lambda$ and $R$ where $\lambda_{cr}$ and
 $R_{cr}$ represent the critical\\
\hspace*{\parindent} values of $\lambda$ and $R$ with which
 the region preserving the chiral symmetry\\
\hspace*{\parindent} is divided from the one of
 the broken chiral symmetry.
\vglue 0.8cm

\begin{figure}
\end{figure}


\begin{thebibliography}{8}

   \bibitem{EF}    For references see, e. g. ,
                   I.~L.~Buchbinder, S.~D.~Odintsov and I.~L.~Shapiro,
                   {\it Effective Action in Quantum Gravity}
                   (IOP Publishing, Bristol and Philadelphia, 1992).


   \bibitem{RG}    C.~Hill and D.~S.~Salopek,
                   {\it Ann. of Phys.} {\bf 213} (1992) 21 ;\\
                   T.~Muta and S.~D.~Odintsov,
                   {\it Mod. Phys. Lett.} {\bf A6} (1991) 3641.


   \bibitem{NJL}   Y.~Nambu and G.~Jona-Lasinio,
                   {\it Phys.~Rev.} {\bf 122} (1961) 345.


   \bibitem{GR}    C.~W.~Misner, K.~S.~Thorne and J.~A.~Wheeler,
                   {\it Gravitation}
                   (W. H. Freeman and Co., San Francisco, 1973).


   \bibitem{SPT}   J. Schwinger, {\it Phys.~Rev.} {\bf 82} (1951) 664.


   \bibitem{RNC}   See, e. g. [2].


   \bibitem{PT}    L.~Parker and D.~J.~Toms,
                   {\it Phys. Rev.} {\bf D29} (1984) 1584.


   \bibitem{SH}    G.~M.~Shore,
                   {\it Ann. of Phys.} {\bf 128} (1980) 376;\\
                   B.~Allen,
                   {\it Nucl. Phys.} {\bf B226} (1983) 228;\\
                   K.~Ishikawa,
                   {\it Phys.~Rev.} {\bf D28} (1983) 2445 ;\\
                   B.~L.~Hu and D.~J.~O'Connor,
                   {\it Phys.~Rev.} {\bf D30} (1984) 743.


   \bibitem{GN}    D.~J.~Gross and A.~Neveu,
                   {\it Phys. Rev.} {\bf D10} (1974) 3235.



   \bibitem{GNC}   H.~Itoyama,
                   {\it Prog. Theor. Phys.} {\bf 64} (1980) 1886 ;\\
                   I.~L.~Buchbinder and E.~N.~Kirillova,
                   {\it Int. J. of Mod. Phys.} {\bf A4} (1989) 143.



\end{thebibliography}
\end{document}